\documentclass[english,two column]{revtex4-1}
\usepackage[T1]{fontenc}
\usepackage[latin9]{inputenc}
\usepackage{bm}
\usepackage{color}
\usepackage{amsmath}
\usepackage{amssymb}
\usepackage{graphicx}
\usepackage{babel}
\usepackage{polski}

\begin{document}

\preprint{APS/123-QED}

\title{Finding the stable mechanism of ring solitons in two-dimensional Fermi superfluids}
\author{Hao-Xuan Sun$^1$}
\author{Liu-Yang Cheng$^2$}

\author{Shi-Guo Peng$^{2,3,4}$}
\email{pengshiguo@wipm.ac.cn}
\author{Yan-Qiang Li$^1$}
\email{lyq_qdsd@163.com}
\author{Peng Zou$^1$}
\email{phy.zoupeng@gmail.com}
\affiliation{$^1$Centre for Theoretical and Computational Physics, College of Physics, Qingdao University, Qingdao 266071, China}
\affiliation{$^2$State Key Laboratory of Magnetic Resonance and Atomic and Molecular Physics, Innovation Academy for Precision Measurement Science and Technology, Chinese Academy of Sciences, Wuhan 430071, China}
\affiliation{$^3$Center for Theoretical Physics, Hainan University, Haikou 570228, China}
\affiliation{$^4$School of Physics and Optoelectronic Engineering, Hainan University, Haikou 570228, China}

\begin{abstract}
We theoretically investigate the stable mechanism of a ring soliton in two-dimensional Fermi superfluids by solving the Bogoliubov-de Gennes equations and their time-dependent counterparts. In the uniform situation, we discover that the ring soliton is always driven away from its initial location, and moves towards the boundary due to a curvature-induced effective potential. The ring soliton is impossible to remain static at any location in the uniform system. To balance the density difference between the ring soliton's two sides, a harmonic trap is introduced, which can exert an effect to counterbalance the curvature-induced effective potential. This enables the ring dark soliton to become a stable state at a particular equilibrium position $r_s$, where the free energy of the ring dark soliton just reaches the maximum value. Once the ring soliton is slightly deviated from $r_s$, some stable periodic oscillations of the ring soliton around $r_s$ will occur. Some dissipation will occur to ring soliton once its minimum radius is comparable to the healing length of soliton's Friedel oscillation. This dissipation will increase the oscillation amplitude and finally make the ring soliton decay into sound ripples. Our research lays the groundwork for a more in-depth understanding of the stable mechanism of a ring dark soliton in the future.
\end{abstract}


\maketitle

\section{\label{sec:introduction} Introduction }

As a spatially localized wave, the soliton originates from the interplay between the dispersion and non-linearity of the underlying system \cite{DrazinBook2002}. Usually, the dark soliton is a kind of low-energy solution of many-body systems and exerts a crucial influence on the dynamical behavior of the system. In the past three decades, much research attention has focused on solitons of atomic Bose-Einstein condensates (BECs) \cite{KevrekidisBook2007}. Many members of the soliton family are investigated in different physical systems (or matter states), including bright solitons in attractive BECs and dark solitons in repulsively interacting BECs \cite{DenschlagSCI2000,BurgerPRL1999,AndersonPRL2001,SakaguchiJPB2004,StellmerNP2008}. Similar research has also been carried out in Fermi superfluids, to study the static dark soliton and the dynamical motion behavior of solitons \cite{AntezzaPRA2007,LiaoPRA2011,ScottPRL2011}. The decay mechanism of solitons, known as snake instability, has also been theoretically investigated \cite{CetoliPRA2013}. Since the realization of the spin-orbit-coupling (SOC) effect in ultracold
atomic gases, a novel Majorana soliton and its distinct motion mechanism have been discovered in SOC Fermi superfluids \cite{XuPRL2014,LiuPRA2015,MateoPRA2022,ZouPRL2016}. The soliton can be generated by a phase-imprinting technique \cite{SachaPRA2014,KongPRA2021}. Moreover, it has been experimentally fabricated to observe its dynamical behavior \cite{YefsahNAT2013}. 

Different from other solitary waves with a long life, the typical characteristic of a dark soliton is that there is a phase jump $\delta\phi=\pi$ exhibited by the phase of the order parameter across the two sides of the soliton's position. The typical soliton discussed above is a linear one, in which the value of the phase jump $\delta\phi$ changes along an ideal straight line. This also marks the location of the soliton, where the core density is usually lower than the bulk density of the system. The soliton also displays some fluctuations in both density and order parameter within the regime of healing length, which is called Friedel oscillation \cite{AntezzaPRA2007,QuJLTP2021}. If certain perturbations disrupt the straight-line geometry of linear solitons, the snake instability will cause the soliton to bend and possibly decay into sound waves or vortices, etc \cite{CetoliPRA2013}. Generally, it seems that all the bending or disturbing behavior in the perpendicular direction of the soliton will break its stability. However, this is not always true. Besides the linear soliton, it is of great interest to explore the existence of other geometries of stable solitons. In fact, ring solitons have also been investigated in BECs. The geometric locus of its phase jump forms an ideal circular shape \cite{TheocharisPRL2003,WangPRA2019}. However, this ring soliton is usually not a stable state. It often tends to exhibit some extraordinary dynamical behaviors and evolve into various forms, such as multiple concentric ring solitons, vortex-antivortex pairs, or a complex pattern structure \cite{TheocharisPRL2003,GuoAPS2020,WangPRA2021,TamuraPRX2023}. It is interesting to understand and control these rich dynamical behaviors, which are usually not easy to carry out.
In Fermi superfluid, there are quite a few discussions about the ring soliton currently. Reference \cite{BarkmanPRR2020} introduces stable ring solitons in an imbalanced fermionic system with a phenomenological Ginzburg-Landau theory. Sometimes, the ring soliton is referred to as "spin-polarized droplets" or "ferrons" in the imbalanced Fermi superfluid~\cite{MagierskiPRA2019,TuzemenAPPB2020,MagierskiPRA2021,TuzemenNJP2023}. In this paper, we will study the ring soliton in balanced two-dimensional (2D) Fermi superfluids with both static and time-dependent Bogoliubov-de Gennes (BdG) equations. The system is confined in a circular disc potential, which has been realized in experiment~\cite{ChomazNC2015,MukherjeePRL2017,HueckPRL2018,NavonNP2021}.The purpose of this paper is to better understand the motion and decay mechanism of ring solitons and find a possible physical explanation behind them.

This paper is organized as follows. In the subsequent section, we will first introduce the static BdG equations and their time-dependent counterparts, by which to examine the stability of ring solitons. In Sec.~\ref{sec:results}, we use four subsections to investigate the dynamical behavior of ring solitons in both uniform and trapped systems, to analyze the stability and decay behavior of ring solitons. Finally, our conclusions are given in Sec.~\ref{sec:cons}.

\section{\label{sec:methods} Methods}
We consider a 2D Fermi superfluid with balanced populations of the two spin components at zero temperature. Particles of different spins will experience an s-wave contact interaction. Within the mean-field theory, the many-body Schr\"odinger equations of this system can be well reduced to BdG equations 
\begin{equation}
\begin{aligned}
 \begin{bmatrix}
 \hat{H}_S & \Delta(\textit{\textbf{r}}) \\ \Delta^*(\textit{\textbf{r}}) & - \hat{H}_S
 \end{bmatrix}
 \begin{bmatrix}
 u_\eta(\textit{\textbf{r}}) \\ v_\eta(\textit{\textbf{r}})
 \end{bmatrix}
 =E_\eta 
 \begin{bmatrix}
 u_\eta(\textit{\textbf{r}}) \\ v_\eta(\textit{\textbf{r}})
 \end{bmatrix},\label{eq:static_BdG}
\end{aligned}
\end{equation}
where $\hat{H}_S=-\nabla^2/2m+V(\textit{\textbf{r}})-\mu$ is the single particle Hamiltonian in an external potential $V(\textit{\textbf{r}})$ with mass $m$ and chemical potential $\mu$. $E_\eta$ is the quasiparticle eigenenergy with the corresponding quasiparticle wave functions $u_\eta$ and $v_\eta$, which are normalized by $\int d\textit{\textbf{r}}[v_\eta(\textit{\textbf{r}})v^*_{\eta'}(\textit{\textbf{r}})+u_\eta(\textit{\textbf{r}})u^*_{\eta'}(\textit{\textbf{r}})]=\delta_{\eta\eta'}$. Here and hereinafter, we set the Planck constant $\hbar=1$. These BdG equations should be self-consistently solved with the order parameter (or gap) function 
\begin{equation}
\begin{aligned}
\Delta(\textit{\textbf{r}})=g\sum_\eta u_\eta(\textit{\textbf{r}}) v_\eta^*(\textit{\textbf{r}})
\end{aligned}
\label{eq:gap}
\end{equation}
 and total density function 
 \begin{equation}
\begin{aligned}
n(\textit{\textbf{r}})=2\sum_\eta |v_\eta(\textit{\textbf{r}})|^2.
\end{aligned}
\label{eq:den}
\end{equation}
Here $g$ is the bare interatomic interaction strength, and it should be regularized by 
\begin{equation}
    \begin{aligned}
        \frac{1}{g}=-\sum_k\left(\frac{1}{E_b+2\epsilon_k}\right),
    \end{aligned}
\end{equation}
in which $\epsilon_k=k^2/2m$ and binding energy $E_b$ describes the strength of interaction. A cutoff energy $E_c$ is considered during the summation of possible energy levels in the numerical calculation, namely $0<E_\eta<E_c$. To investigate the dynamics of the system, time-dependent BdG equations, whose expressions are 
\begin{equation}
\begin{aligned}
 \begin{bmatrix}
 \hat{H}_S & \Delta(\textit{\textbf{r}},t) \\ \Delta^*(\textit{\textbf{r}},t) & - \hat{H}_S
 \end{bmatrix}
 \begin{bmatrix}
 u_\eta(\textit{\textbf{r}},t) \\ v_\eta(\textit{\textbf{r}},t)
 \end{bmatrix}
 =i\frac{\partial}{\partial t}
 \begin{bmatrix}
 u_\eta(\textit{\textbf{r}},t) \\ v_\eta(\textit{\textbf{r}},t)
 \end{bmatrix},\label{eq:td_BdG}
\end{aligned}
\end{equation} 
are utilized to observe the time evolution of any initial state. 
The free energy of the system is 
\begin{equation}
    F=\int d\textit{\textbf{r}}\sum_\eta\left[2\left(\mu-E_\eta\right)|v_\eta|^2+\Delta^*u_\eta v_\eta^*\right]+\mu N,
    \label{eq:free}
\end{equation}
where $N=\int d\textit{\textbf{r}}n(\textit{\textbf{r}})$ is the total particle number of the system. In the following, we will consider both the uniform system ($V(\textit{\textbf{r}})=0$) and the trapped one ($V(\textit{\textbf{r}}) \neq 0$). These two cases have their own typical wave vector $k_F$ and typical energy $E_F=k_F^2/2m$, which are used to carry out dimensionless treatment of all related physical quantities in the following discussion.

 Owing to the geometric symmetry of the ring soliton, we study the system with a 2D circular hard-wall environment with a maximum radius $R$. Naturally, it is better to introduce the position parameter $\textit{\textbf{r}}=(r,\theta)$ in polar coordinates instead of $\textit{\textbf{r}}=(x,y)$ in Cartesian coordinates. In this case, the quasiparticle wave functions $u_\eta$ and $v_\eta$ could be expanded with a set of normalized basis functions $\psi_{jl}(r)=\sqrt{2}J_l(r\alpha_{jl}/R)/RJ_{l+1}(\alpha_{jl})$, where $\alpha_{jl}$ is the $j$th zero of the Bessel function $J_{l}(r)$. Namely $u_\eta(r,\theta)=\sum_j c_{\eta j}\psi_{jl}(r)e^{il\theta}/\sqrt{2\pi}$ and $v_\eta(r,\theta)=\sum_j d_{\eta j}\psi_{jl}(r)e^{il\theta}/\sqrt{2\pi}$. Then the process of solving BdG Eqs.~\ref{eq:static_BdG} is reduced to a mathematical matrix diagonalization problem with
\begin{equation}
\begin{aligned}
 \begin{bmatrix}
 H^l_{jj'} & \Delta^l_{jj'} \\ \Delta^{l}_{j'j*} &  -H^{l}_{jj'}
 \end{bmatrix}
 \begin{bmatrix}
 c_{\eta j} \\  d_{\eta j}
 \end{bmatrix}
 =E_\eta\begin{bmatrix}
 c_{\eta j} \\  d_{\eta j}
 \end{bmatrix},\label{eq:BdG_basis}
\end{aligned}
\end{equation}
in which $H^l_{jj'}=(a^2_{jl}/R^2-\mu)\delta_{jj'}+V^l_{jj'}$ and $\Delta^m_{jj'}=\int \phi_{jl}(r)\Delta(r)\phi_{j'l}^*(r)rdr$. In our numerical calculation, we have set parameters $E_c=25E_F$ with the maximum angular quantum number $|l|_{max}=130$ and the zero-point number $j_{max}=60$, whose values we have checked are large enough to carry out numerical calculations.

\section{\label{sec:results}results of the ring soliton} 

\subsection{Uniform system}
\begin{figure}
    \centering
\includegraphics[width=1.0\linewidth]{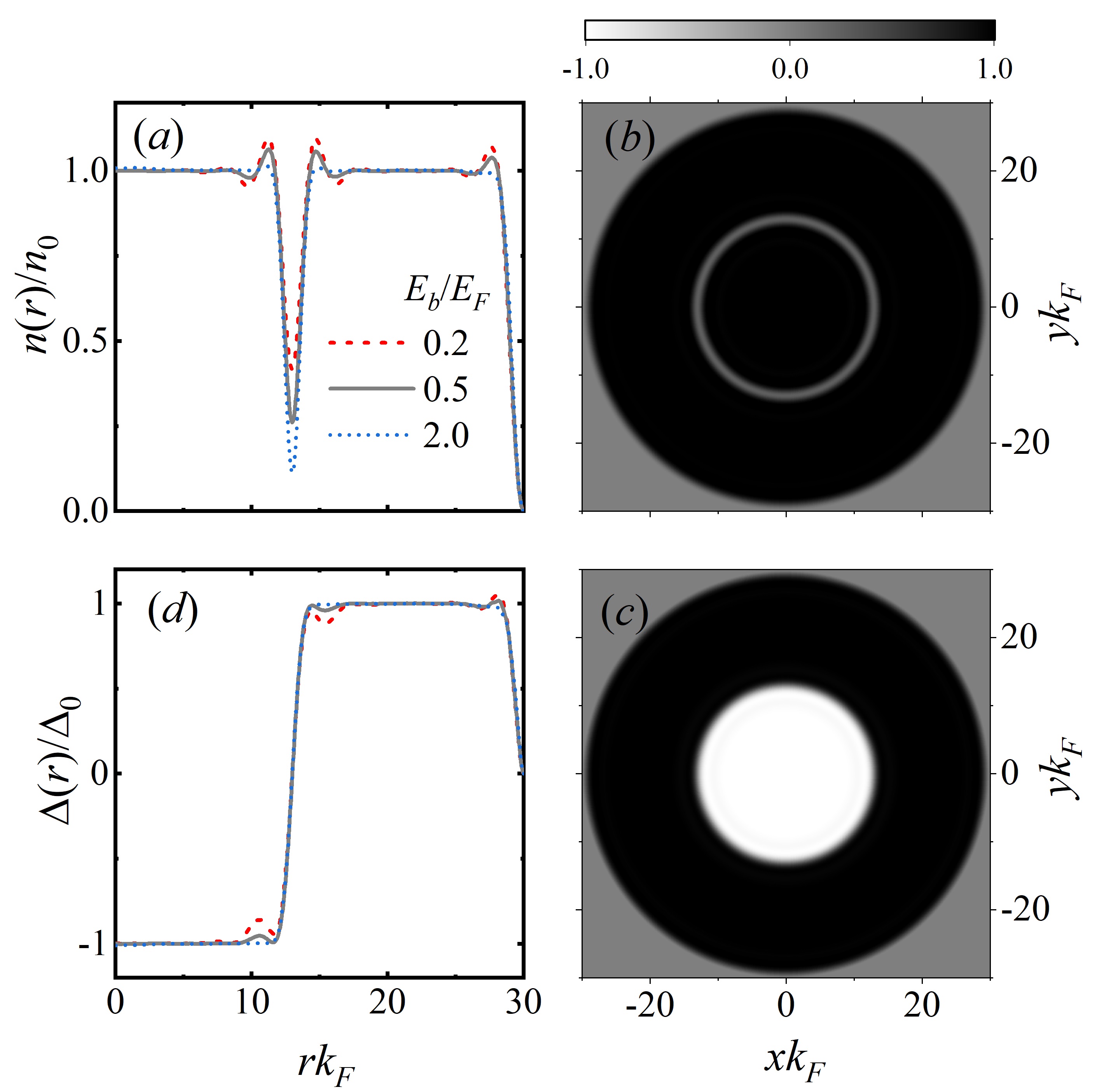}
    \caption{The density (upper panels) and order parameter (lower panels) distributions of an instantaneous ring dark soliton in both the radial direction (left panels) and $x$-$y$ plane (right panels). The soliton is located at located at $r_0k_F=13.0$. The binding energy $E_b=0.5E_F$.}
    \label{Fig:rds}
\end{figure}

The uniform system has a well-defined bulk density $n_0$, by which the typical wave vector $k_F=\sqrt{2\pi n_0}$.
Generally all possible many-body solutions of a fermionic system can be derived by self-consistently solving the BdG equations shown in Eq.~\ref{eq:static_BdG}, the order parameter equation presented in Eq.~\ref{eq:gap}, and the density equation given in Eq.~\ref{eq:den}. In this numerical process, it is necessary to iterate over all possible degrees of freedom to achieve a stable state. For example, one needs to iterate both the chemical potential $\mu$ and the spatial distribution of the order parameter $\Delta(\textit{\textbf{r}})$ in order to obtain a stable linear dark soliton in the uniform system. In the case of the ring soliton, its radius plays the role of another new degree of freedom. However, there is no means for us to obtain an equation for the exact radius. To overcome this problem, our strategy is to consider temporarily the ring soliton in a uniform system by iterating solely the chemical potential $\mu$ and order parameter distribution $\Delta(\textit{\textbf{r}})$ to obtain an instantaneous zero-velocity state. Subsequently, a time-dependent simulation of this instantaneous state is conducted to verify its stability.

Similar to the case of obtaining a linear dark soliton, we first guess an initial radius distribution of the order parameter $\Delta(r)$ with a phase jump $\delta\phi=\pi$ at a certain location $r_0$, whose value is far from both the center and the edge to avoid their influence. By solving self-consistently Eqs.~\ref{eq:static_BdG}, \ref{eq:gap} and \ref{eq:den}, we obtain an instantaneous zero-velocity solution of the ring dark soliton at $r_0k_F=13.0$. As depicted in Fig.~\ref{Fig:rds}, the upper two panels illustrate the spatial distribution of density, while the lower two panels exhibit the spatial distribution of the order parameter. The results of ring solitons in different binding energies $E_b$ are also displayed in the left two panels. As the interaction strength $E_b$ becomes increasingly large, the location of the ring dark soliton presents a weaker and weaker Friedel oscillation in both density (panel (a)) and order parameter (panel (d)). The ring dark soliton also takes on a deeper and deeper density valley, as shown by different line styles in panel (a), which makes soliton understandable as a negative-mass wave.

\begin{figure}
    \centering
    \includegraphics[width=1.0\linewidth]{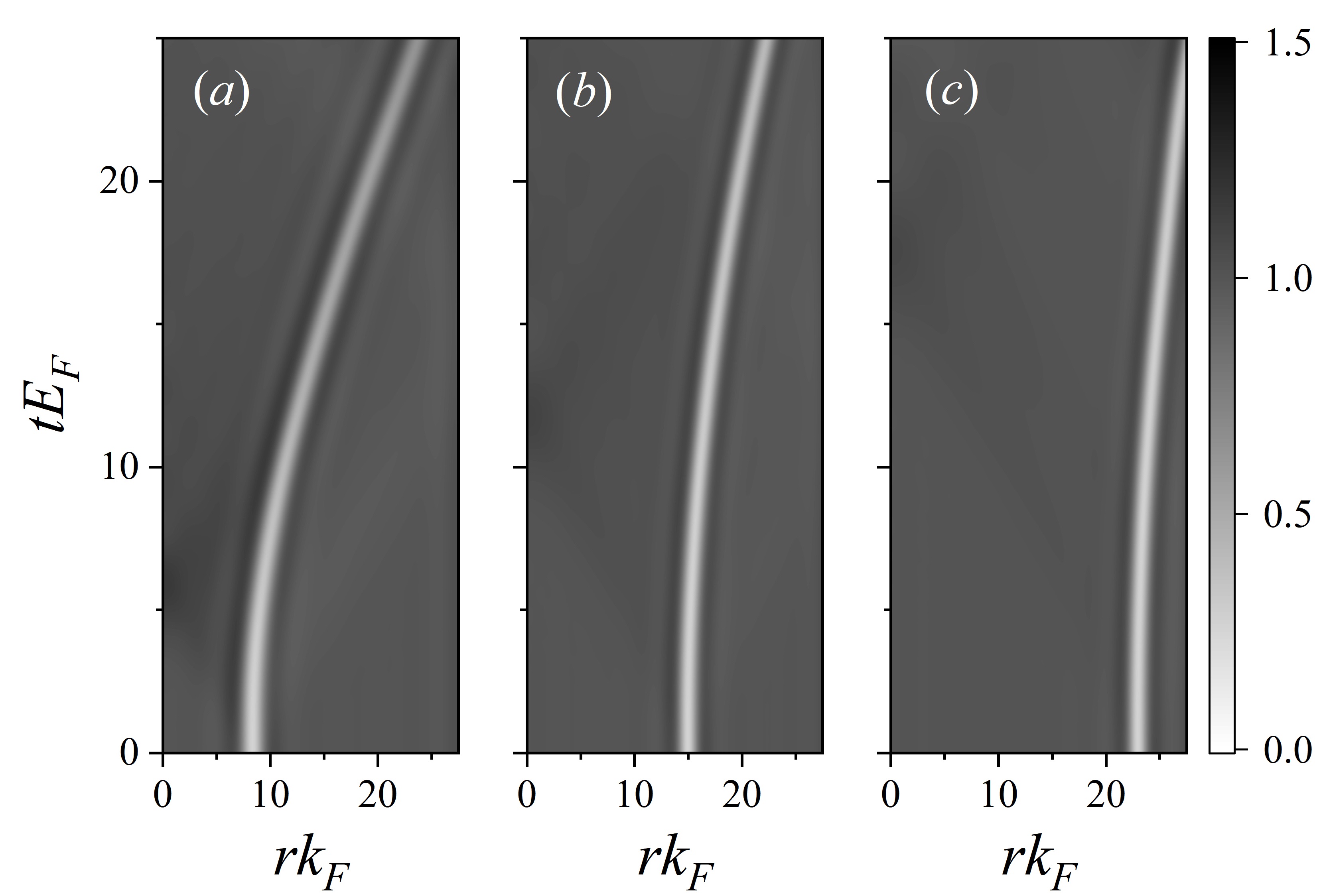}
    \caption{The time evolution of ring solitons in a uniform system with different initial locations, namely (\textit{a}) $r_0k_F=7$, (\textit{b}) $r_0k_F=15$ and (\textit{c}) $r_0k_F=23$. The binding energy $E_b=0.5E_F$. }
    \label{Fig:rs_uniform}
\end{figure}

To check the stability of the ring dark soliton, a time-dependent simulation is the most convenient method. A stable solution should exhibit a result that does not change over time. We utilize the above instantaneous state as an input one at time $t=0$, and then evolve it using the time-dependent BdG Eqs.~\ref{eq:td_BdG}. 
In some sense, this time-dependent approach plays the role of iterating the location of the ring dark soliton. The results are shown in Fig.~\ref{Fig:rs_uniform}. We obtain time evolutions of three ring dark solitons, which are initially located at (a) $r_0k_F=7$, (b) $r_0k_F=15$ and (c) $r_0k_F=23$. Unfortunately, instead of being fixed at their initial position $r_0$ all the time, all three panels demonstrate that ring solitons are driven away from $r_0$, and move consistently outward to the edge, induced by one common motion mechanism. This verifies that the ring solitons in the uniform system are not stable states. It can also be readily observed that the nearer the initial location $r_0$ of a ring dark soliton is to the centre, the farther it moves away from the centre. It should be emphasized that there is no external trap in the uniform system to generate such an orientated motion. 

Physically, this common mechanism arises from the interplay between particles, whose strength should also be related to the geometric shape of ring solitons. The ring soliton is a kind of density-valley structure and has fewer particles than those in the bulk position. The circular geometry of a ring soliton means that there will be more particles on the outer side than on the inside. This kind of density distribution makes the soliton a negative-mass density wave. At the same interaction strength between particles, more particles mean that the outside of the ring soliton will possibly bring relatively more particles than those on the inner side, inducing the valley structure to move outward. In the following, we will call this effect a curvature-induced effective potential, whose analytical expression is temporarily unknown and is absent in the conventional linear soliton due to its symmetric structure. The existence of this very curvature-induced effective potential not only explains the phenomenon in Fig.~\ref{Fig:rs_uniform}, but also makes it impossible for a ring dark soliton to become a stable state in the uniform system. In fact, the above discussions also shine a light on building a stable ring dark soliton in the non-uniform system. 

\begin{figure}
    \centering
    \includegraphics[width=1.0\linewidth]{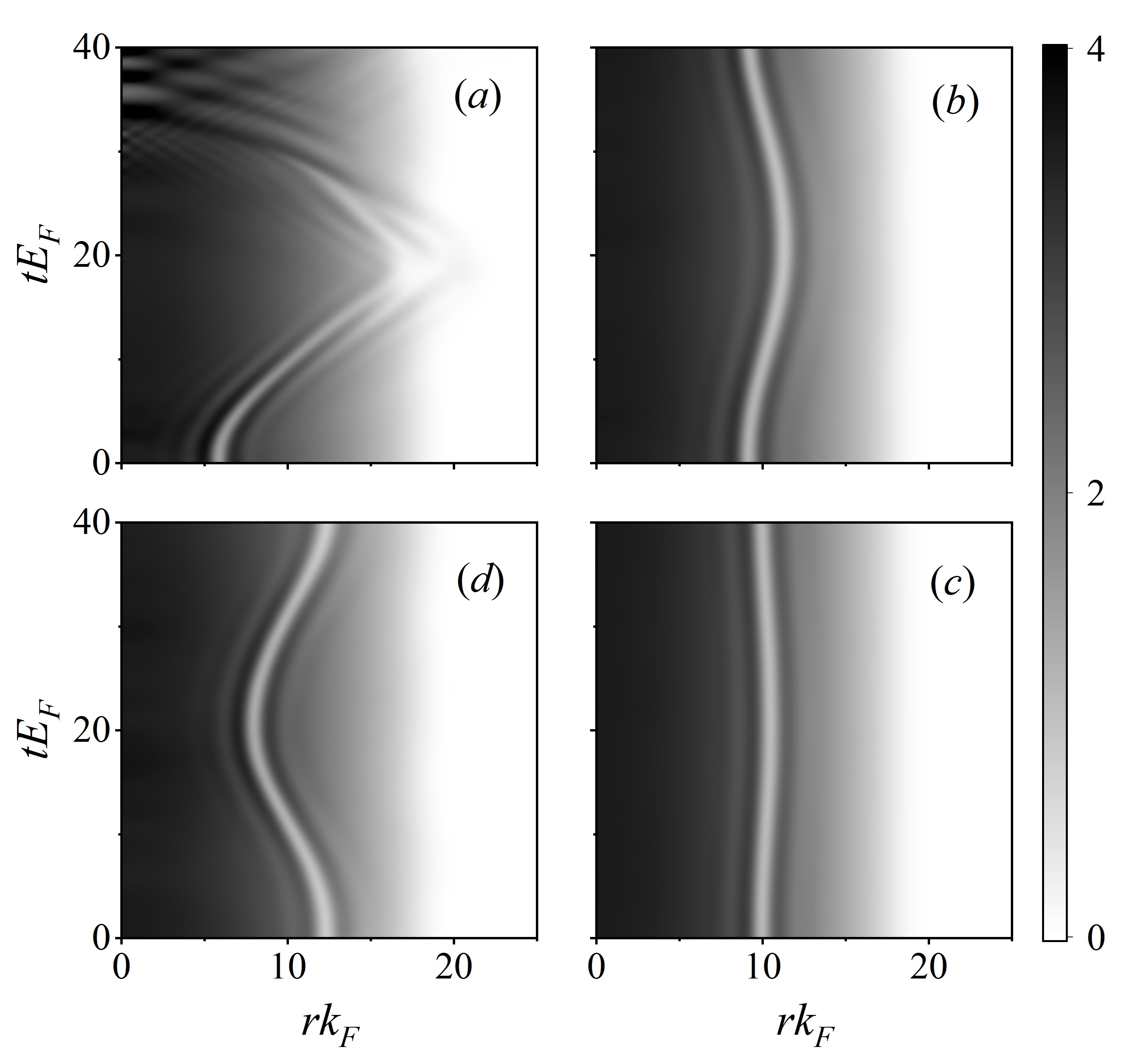}
    \caption{The time evolution of ring solitons in a harmonic trap with different initial locations, namely (\textit{a}) $r_0k_F=5.5$, (\textit{b}) $r_0k_F=9.0$, (\textit{c}) $r_0k_F=10.1$ and (\textit{d}) $r_0k_F=12.2$. The binding energy $E_b=0.5E_F$.}
    \label{Fig:rs_trap}
\end{figure}

\subsection{System in a harmonic trap}

\begin{figure}
    \centering
    \includegraphics[width=0.95\linewidth]{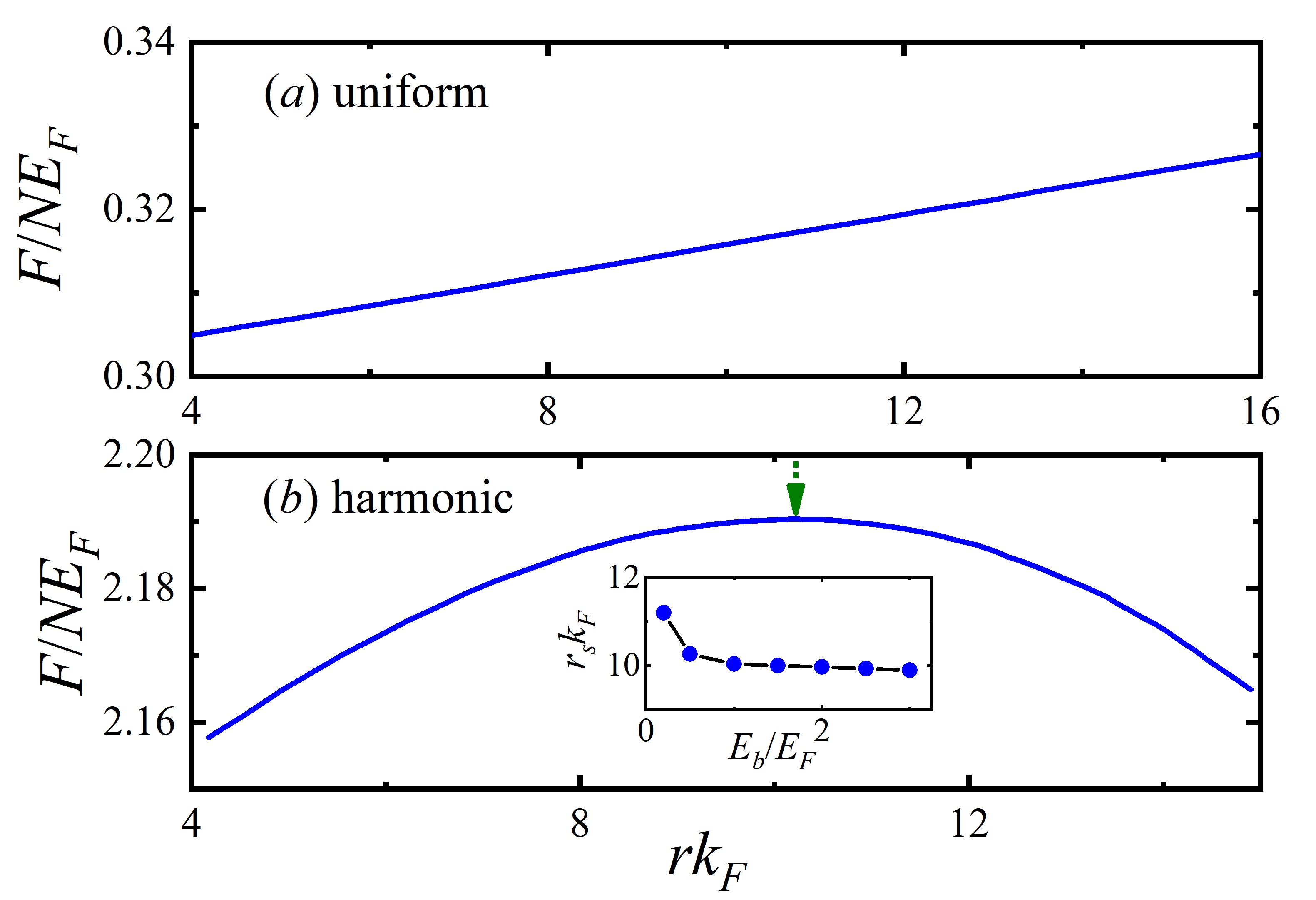}
    \caption{The free energy of instantaneous ring dark solitons at different locations. (a) uniform system with the same parameters as that in Fig.~\ref{Fig:rds} and Fig.~\ref{Fig:rs_uniform}; (b) system in harmonic trap with the same parameters as that in Fig.~\ref{Fig:rs_trap}. The inset figure shows the stable location of ring soliton at different binding energy $E_b$.}
    \label{Fig:free}
\end{figure}

By reducing the outer density of ring solitons while increasing their inner density, it is possible to counteract their tendency to move outward. An intuitive strategy to generate this effect involves employing a harmonic trap $V(\textbf{\textit{r}})=m\omega^2r^2/2$, which is frequently utilized in experiments. Of course, the harmonic trap is not the only candidate for an external potential. In this part, we will consider the influence of a harmonic trap on the system and set the particle number $N$ of the system as a constant value. In this case, the typical wave vector is $k_F=\sqrt{m\omega}$. The dynamics of a linear soliton in the harmonic trap have been introduced in Ref.~\cite{ScottPRL2011}.The harmonic trap is capable of endowing a linear soliton with an inherent tendency to move inward toward the trap centre. Therefore, it is interesting to explore the competition between the harmonic trap and curvature-induced effective potential, as well as their impacts on the stability of the ring dark soliton. To date, it remains unclear which of the two will prevail over the other, or if they will end up evenly matched.

To investigate the underlying mechanism, we continue to employ a time-dependent simulation approach to study the competition process between these two motion trends. Analogous to the approach used in the uniform case, we generate instantaneous ring dark solitons at four different initial locations. The results are shown in
Fig.~\ref{Fig:rs_trap}. From panel (a) to panel (d), instantaneous ring dark soliton is initially located at $r_0k_F=5.5,9.0,10.0$ and $12.2$, respectively. During the progress of time evolution, the ring soliton in panel (a) moves outward, which is similar to the uniform case and indicates that the curvature-induced effective potential dominates over the effect of the harmonic trap. The ring soliton exhibits a significantly large oscillation amplitude. However, when $r_0$ is increasingly far away from the centre, for example $r_0k_F=9.0$ in panel (b), the ring soliton displays a narrow oscillation amplitude, although it still moves outward initially. Interestingly, in panel (c) where $r_0k_F=10.0$, it seems that the ring soliton is trying to be static at this position and shows an almost unobservable oscillation. Across this critical position, for instance $r_0k_F=12.2$ in panel (d), the ring soliton initially moves inwards and again takes on an obvious oscillation amplitude. All these panels confirm that the outcome of the competition between the harmonic trap and the curvature-induced potential is location-dependent. Notably, there exists a critical location at which the contributions of the aforementioned two factors are balanced with each other. This makes the ring dark soliton a stable state at this equilibrium position ($r_sk_F\approx 10.0$).   

To understand the underlying physical reason for generating stable ring soliton at $r_s$, it is better to calculate the free energy of the instantaneous ring dark soliton. As shown in panel (b) of Fig.~\ref{Fig:free}, the free energy takes on a local maximum around $r_sk_F= 10.1$, which is marked by a dotted arrow. This location is almost the same as one in panel (c) of Fig.~\ref{Fig:rs_trap}. Usually, a stable state of a many-body system should appear at the local (or global) minimum of free energy. In fact, this is true for the conventional wave of particles with positive mass. However, it is opposite for the soliton with negative mass. Thus, it can be understood physically why the stable soliton is located where free energy reaches its maximum. Similar work on the free energy of the uniform ring dark soliton has also been conducted in panel (a), which shows monotonically linear behavior without any local maximum or minimum. This also helps to explain why there is no stable ring dark soliton state in the uniform system. Here we use the outcome of the competition between the harmonic trap and curvature-induced effective potential to study the stable mechanism of the ring soliton. In fact, this strategy is universal and can be extended to various other complex systems characterized by the presence of ring dark solitons, including the imbalanced Fermi superfluid in Ref.~\cite{BarkmanPRR2020}. 

\begin{figure}
    \centering
    \includegraphics[width=1.0\linewidth]{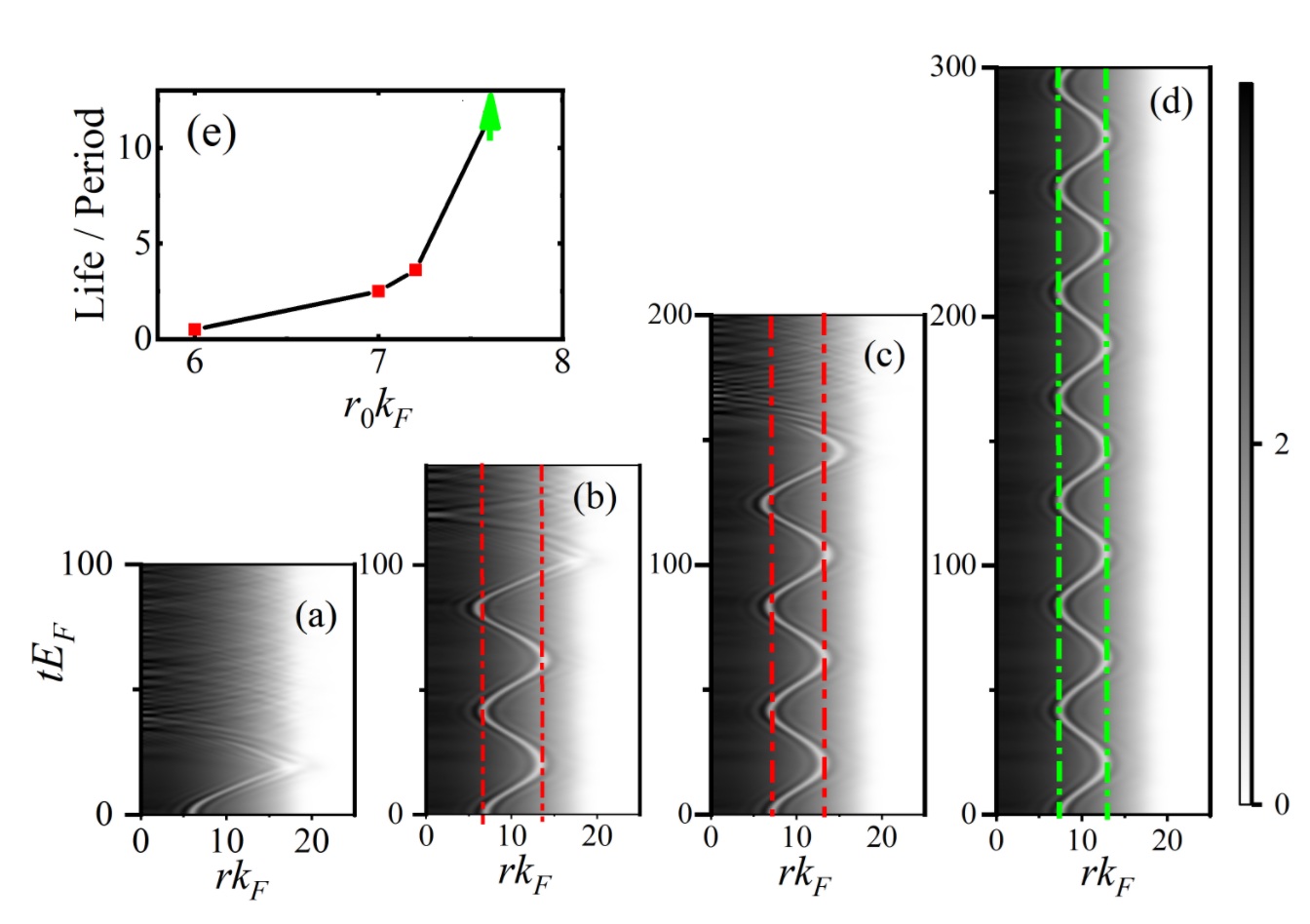}
    \caption{The time-evolution of ring solitons whose initial positions $r_0$ locate on the left-hand side of the stable location $r_sk_F=10.1$. Their initial positions are respectively (a) $r_0k_F=6.0$, (b) $r_0k_F=7.0$, (c) $r_0k_F=7.2$ and (d) $r_0k_F=7.6$. The binding energy $E_b=0.5E_F$. Panel (e) shows the ratio between approximation life and oscillation period of ring soliton at previous four different positions. The green arrow here marks the maximum time ($tE_F=500$) we run in our time simulation when ring soliton is still stable. One oscillation period is around $42/E_F$.}
    \label{Fig:left}
\end{figure}

We also numerically calculate the equilibrium location $r_s$ at different binding energies $E_b$. As shown by the inset figure in panel (b) of Fig.~\ref{Fig:free}, we can always obtain an equilibrium location $r_s$, whose value decreases with $E_b$ and finally
appears to be a constant value.

\subsection{Decay of ring solitons}

\begin{figure}
    \centering
    \includegraphics[width=1.0\linewidth]{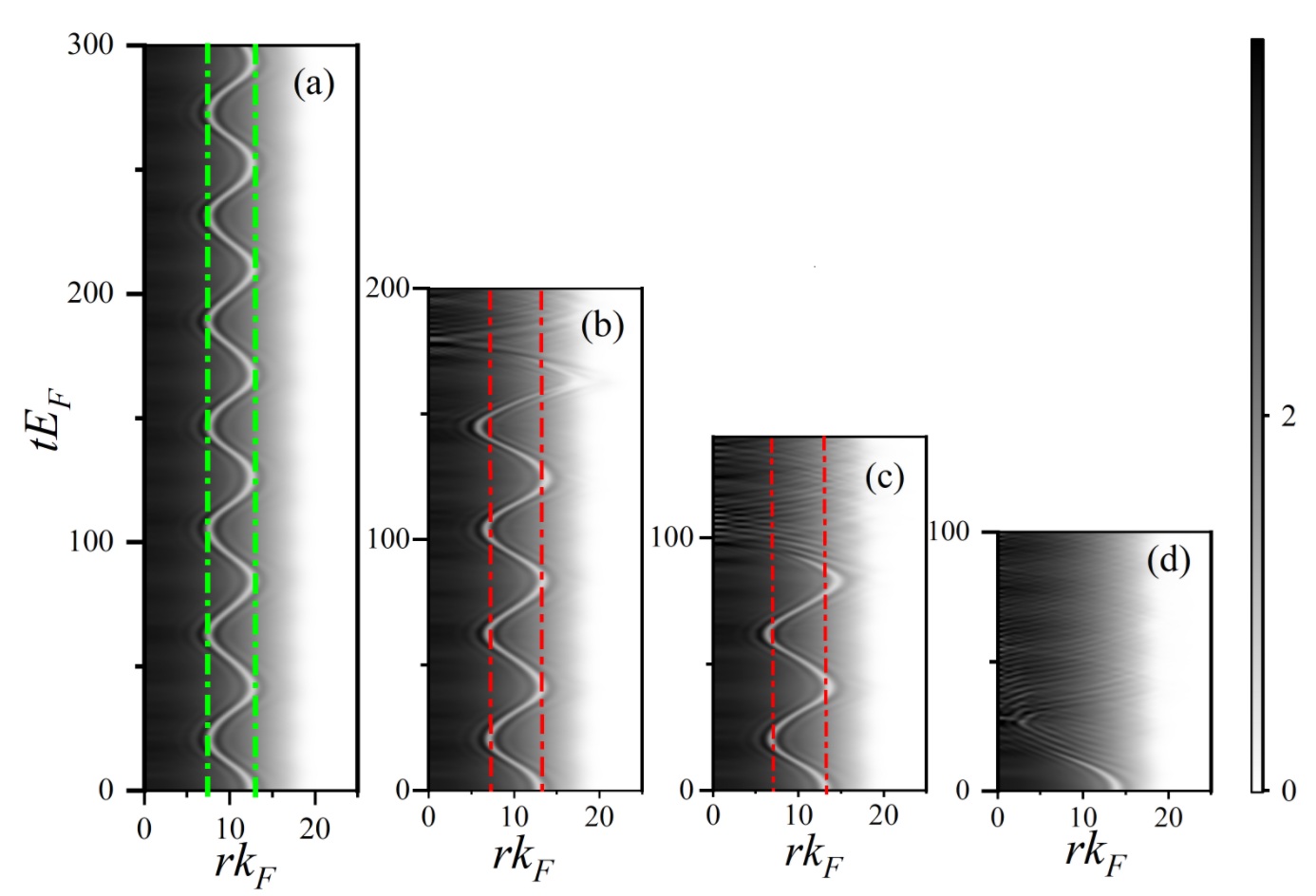}
    \caption{The time-evolution of ring solitons whose initial positions $r_0$ locate on the right-hand side of the stable location $r_sk_F=10.1$. Their initial positions are respectively (a) $r_0k_F=12.6$, (b) $r_0k_F=13.0$, (c) $r_0k_F=13.2$ and (d) $r_0k_F=14.0$. The binding energy $E_b=0.5E_F$.}
    \label{Fig:right}
\end{figure}

Next, it is natural to investigate what will happen when the ring dark soliton is not exactly located at the equilibrium position $r_s$. It will be trivial if some dissipation immediately occurs to the ring soliton when its location is slightly deviated from $r_s$, and it will cause the ring soliton to decay. To check this, we carry out several time evolutions of the ring soliton at different initial locations. As shown in Fig.~\ref{Fig:left}, we prepare four zero-velocity ring solitons in different initial locations, namely $r_0k_F=6.0,7.0,7.2$ and $7.6$. To our surprise, we find different kinds of dynamical oscillation behaviors of ring solitons. In panel (a), the initial radius $r_0$ of the ring solitons is almost of the order of its healing length displayed by the Friedel oscillation. The ring soliton quickly decays into sound wave ripples in a short time. The dynamical behaviors in panel (b) and (c)  are different from those in panel (a), we first observe a few periodic oscillations of ring solitons around the equilibrium position $r_s$. The oscillation period is around $42/E_F$, and its value is determined by some inner parameters of the system \cite{ScottPRL2011}. It seems that a dissipation phenomenon occurs in ring solitons, which gradually increases their oscillation amplitude. This phenomenon could be clearly observed by some red dot-dashed vertical lines, which initially marks the location with the maximum oscillation amplitude. Finally, the ring soliton in panel (b) totally decays into sound ripples in its third oscillation period, while the one in panel (c) in its fourth period due to its relatively bigger initial position $r_0$ (or closer to $r_s$). Different from other panels, the ring soliton in panel (d) experiences a longer oscillation time with a constant oscillation amplitude (marked by green dot-dashed lines). We carry out the simulation for a long time ($tE_F=300$), whose value is nearly twice as large as that in panel (c), while its $r_0$ is just slightly larger than that in panel (c). No significant variation in its oscillation amplitude is observed, and no sound ripples are generated. We have checked that the ring soliton in panel (d) is still stable at time $tE_F=500$ (the green arrow in panel(e)), the life of ring soliton should be much longer and thus impossible to attain within the simulation duration. The rapidly increasing trend in the lifetime of the ring soliton in panel (e) indicates the existence of a critical oscillation radius, beyond which no dissipation occurs to the ring soliton during its periodic oscillation. It is of interest to explore the physical origins of these distinct dynamical phenomena.

It is obvious that the existence of ring solitons requires some space to present a non-zero radius. We find that these different dynamical behaviors can be explained by the competition between the ring soliton's Friedel oscillation and its periodic oscillation around $r_s$. This competition is strong when $r_0$ comes into the effective regime of Friedel oscillation, bringing energy dissipation to the periodic oscillation of the ring soliton, and increasing its maximum velocity (or oscillation amplitude). Once the maximum velocity reaches a critical velocity, determined by the minimum value between sound velocity and pair-breaking velocity, the ring soliton will completely decay into sound ripples \cite{ScottNJP2012}. This competition is absent in the periodic oscillation of linear solitons due to their straight-line geometry \cite{ScottPRL2011}. Once the radius is larger than the healing length of the soliton's Friedel oscillation, this competition becomes weaker and weaker and finally disappears after a sufficiently large radius, in which the ring soliton can always experience a periodic oscillation without any dissipation, just like the one in panel (d). Recently, we have noticed that a similar dynamical behavior also occurs in the ring soliton of the spin-1 BEC system\cite{YuARXIV2025}.

Different from the case in Fig.~\ref{Fig:left}, all initial locations $r_0$ of ring solitons in Fig.~\ref{Fig:right} are on the right-hand side of the equilibrium position $r_s$. Although all their initial radii $r_0$ are much larger than the healing length of Friedel oscillation, this does not mean that their oscillation behaviors are always those without dissipation. In fact, the periodic oscillation of the ring soliton makes it possible to move to the left-hand side of $r_s$, and possibly induces the competition between Friedel oscillation and periodic oscillation of ring solitons. Thus, similar dynamical oscillations of ring solitons can also be investigated,as demonstrated by the four panels of Fig.~\ref{Fig:right}. These dynamical behaviors share the same explanation discussed above. Compared with Fig.~\ref{Fig:left} and Fig.~\ref{Fig:right}, it is easy to find similar dynamical behavior between the left and right-hand sides for a similar oscillation amplitude around $r_s$. 

\subsection{Ideal trap of ring dark soliton}

\begin{figure}
    \centering
    \includegraphics[width=0.95\linewidth]{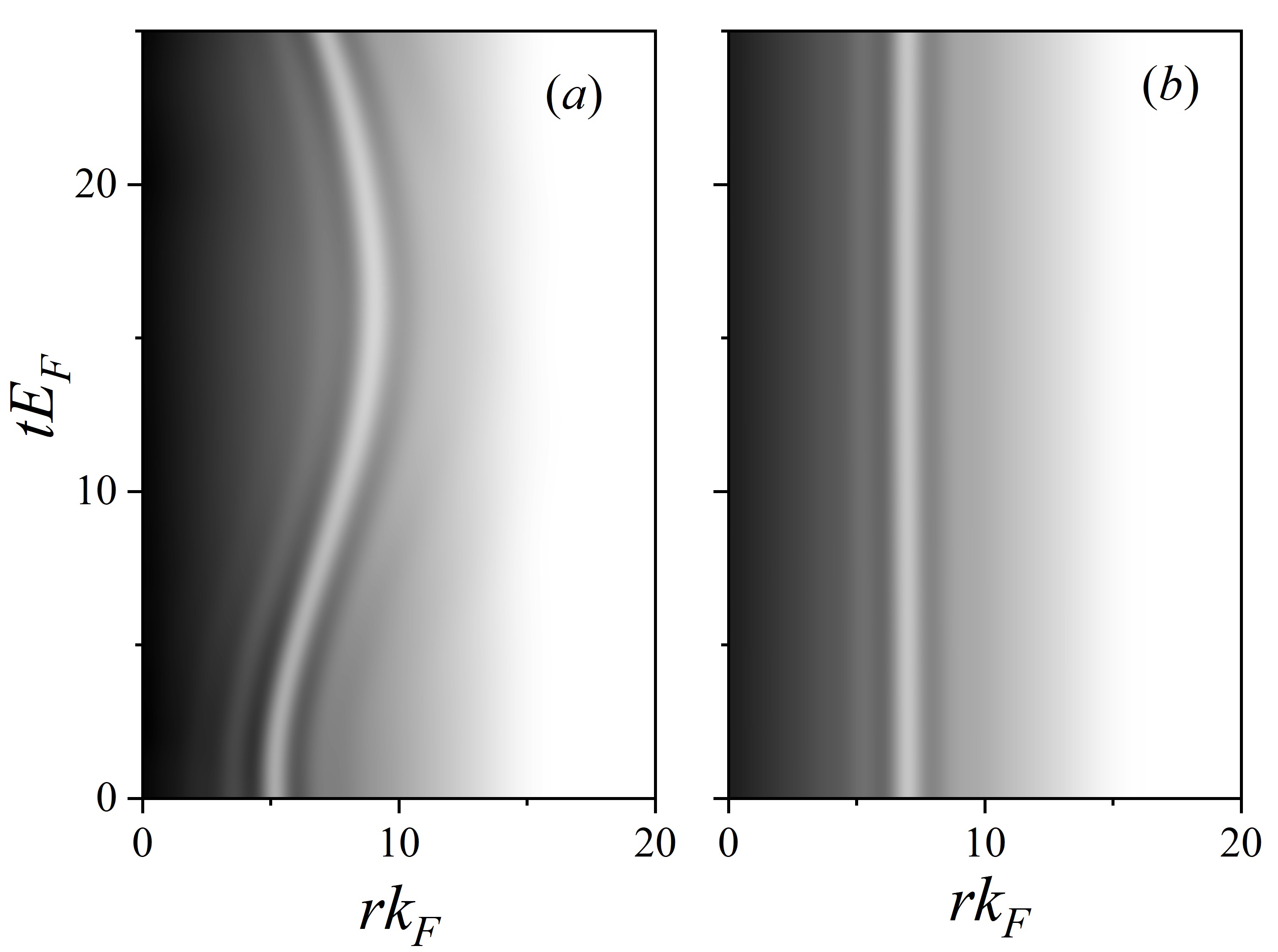}
    \caption{The evolution of ring solitons in a linear trap with different initial location, namely (\textit{a}) $r_0k_F=5$ and (\textit{b}) $r_0k_F=7$. The linear trap satisfies $V(\textbf{\textit{r}})=0.35|r|$. }
    \label{Fig:linear_trap}
\end{figure}

Naturally, an intriguing question emerges: Does there exist an ideal trap capable of stabilizing a ring dark soliton at any position within the external trap? If such a trap exists, what geometric shape should it adopt? From all discussions above, it is clear that the curvature-induced effective potential cannot be well balanced by the harmonic trap except at the location $r_s$. In fact, some key factors may be implied by the free energy in Fig.~\ref{Fig:free}. A linear dependence behavior is shown in panel (a) of Fig.~\ref{Fig:free}. It is easy to remind us to consider a linear external potential, namely $V(\textit{\textbf{r}}) \propto |r|$. We have tried many sets of parameters in this kind of linear trap, and their simulations show similar results. The result of one of them is shown in Fig.~\ref{Fig:linear_trap}, demonstrating that the ring soliton can only be static around $r_0k_F=7$, but not at other locations. This linear trap fails to generate a stable ring dark soliton anywhere. The investigation of an ideal structure of the trap requires the analytical derivation of the expression for curvature-induced effective potential, which is currently unavailable.

\section{\label{sec:cons}CONCLUSIONS}

We theoretically investigate the stable mechanism of the ring soliton in a two-dimensional Fermi superfluid. By solving both static and time-dependent Bogoliubov-de Gennes equations, we find that the competition between the external harmonic trap and the curvature-induced potential plays a key role in building a stable ring dark soliton. In the uniform case, the curvature-induced effective potential always drives the ring soliton away from their initial location, moving outward to the edge, which means there is no stable ring soliton solution in the uniform system. The introduction of a harmonic trap helps balance the curvature-induced effective potential and stabilises the ring dark soliton at an equilibrium location $r_s$, where the free energy of the ring dark soliton just reaches the maximum value. During the periodic oscillation of the ring soliton around $r_s$, we observe another competition between the soliton's Friedel oscillation and the periodic oscillation. This competition is strong once the minimum radius of the ring soliton is on the order of the healing length of Friedel oscillations. A dissipation effect acts on the ring soliton, increasing its oscillation amplitude before ultimately causing it to decay into sound ripples. Once the ring soliton moves sufficiently far from the regime of Friedel oscillations, its periodic oscillations consistently exhibit a fixed amplitude and can persist indefinitely. Our research paves the way for us to understand the stable mechanism of a ring dark soliton in other systems in the future.

\section{ACKNOWLEDGMENTS}
The authors acknowledge fruitful discussions with Xiaoquan Yu and Xiaofei Zhang. This research is supported by the National Natural Science Foundation of China under Grants No. U23A2073 (P.Z.) and No. 12374250 (S.-G.P.), the National Key R$\&$D Program under Grant
No. 2022YFA1404102 (S.-G.P.), and Innovation Program for Quantum Science and Technology under Grant No. 2023ZD0300401 (S.-G.P.).

\end{document}